\documentclass{aa}
\usepackage{graphics}
\usepackage[varg]{txfonts}
\usepackage{natbib}
\usepackage{hyperref}

\newcommand{\dmdt}{\dot{M}}

\newcommand{\zav}[1]{\left(#1\right)}
\newcommand{\hzav}[1]{\left[#1\right]}

\newcommand{\kms}{\ensuremath{\text{km}\,\text{s}^{-1}}}
\newcommand{\vel}{{v}}
\newcommand{\vinfty}{\ensuremath{\vel_\infty}}
\newcommand{\vesc}{\ensuremath{\vel_\text{esc}}}
\newcommand{\ms}{\ensuremath{{M}_{\odot}}}
\newcommand{\msr}{\ensuremath{\ms\,\text{yr}^{-1}}}
\newcommand{\Teff}{\ensuremath{T_\text{eff}}}

\newcommand{\de}{\mathrm{d}}
\newcommand{\teff}{\ensuremath{T_\text{eff}}}

\begin{document}

\title{New line-driven wind mass-loss rates for OB stars with metallicities down
to $0.01\,Z_\odot$}

\author{J.~Krti\v{c}ka\inst{1} \and J.~Kub\'at\inst{2} \and
        I.~Krti\v ckov\'a\inst{1}}

\institute{Department of Theoretical Physics and Astrophysics, Faculty of
           Science, Masaryk University, CZ-611 37 Brno,
           Czech Republic \and
           Astronomical Institute, Czech Academy of Sciences,
           CZ-251 65 Ond\v rejov, Czech Republic }

\date{Received}

\abstract{We provide new line-driven wind models for OB stars with metallicities
down to $0.01\,Z_\odot$. The models were calculated with our global wind code
METUJE, which solves the hydrodynamical equations from nearly hydrostatic
photosphere to supersonically expanding stellar wind together with the equations
of statistical equilibrium and the radiative transfer equation. The models predict
the basic wind parameters, namely, the wind mass-loss rates and terminal
velocities just from the stellar parameters. In general, the wind mass-loss
rates decrease with decreasing metallicity and this relationship steepens for
very low metallicities, $Z\lesssim0.1\,Z_\odot$. Down to metallicities
corresponding to the Magellanic Clouds and even lower, the predicted mass-loss
rates reasonably agree with observational estimates. However, the theoretical
and observational mass-loss rates for very low metallicities exhibit significant
scatter. We show that the scatter of observational values can be caused by
inefficient shock cooling in the stellar wind, which leaves a considerable
fraction of the wind at too high temperatures with waning observational
signatures. The scatter of theoretical predictions is caused by a low number of
lines that effectively accelerate the wind at very low metallicities.}


\maketitle

\section{Introduction}

On the way from their formation toward their final core collapse, massive stars
tend to lose most of their initial mass. The process of mass loss
significantly influences the evolutionary pathways of these stars
\citep{renzovit,kostel}. A considerable fraction of the stellar mass is lost via
line-driven winds. These are long-lasting outflows, which are accelerated mostly
by the momentum acquired from the radiation field via line absorption \citep[for
a review]{vinkara}. However, the implications of winds go far beyond their
impact on the stellar evolution. Winds shape the interaction regions between
massive stars and circumstellar environment \citep{gvarint,kobul}, where the
high-energy particles can be accelerated \citep{bedna,devetatrictvrte}, and
contribute to the enrichment of the interstellar medium with heavy elements
\citep{dorrych}.

Because the stellar winds of hot stars are driven mostly by line transitions of
heavy elements, the strength of the wind depends on the abundance of each
element. The dependence on metallicity is usually parameterized via the total
mass fraction of heavier elements, $Z$, which is often a reasonable proxy. For
instance, in O stars at metallicities higher than about $0.1Z_\odot$, the
rotational mixing, which brings freshly synthesized nitrogen at the expense of
carbon and oxygen to the stellar surface, does not significantly affect basic
wind properties \citep{dusik}.

The mass-loss rate, $\dot M,$ is the key wind parameter, which gives the amount of
mass lost by the star per unit of time. The metallicity dependence of the wind
mass-loss rate is typically parameterized via power law, namely, $\dot M\sim Z^\alpha$.
As follows from theoretical calculations \citep{pusle}, the power-law index,
$\alpha$, is related to the line-strength distribution parameters $\hat\alpha$
and $\hat \delta$ as $\alpha=(1-\hat\alpha)/(\hat\alpha-\hat \delta)$. For a
canonical value $\hat\alpha=2/3$ with $\hat \delta =0$ this gives a square-root
dependence of mass-loss rate on metallicity $\alpha=0.5$. Calculations based on
the Sobolev approximation with a realistic line-list give $\alpha=0.94$
\citep{abpar}, which is not too far but somewhat higher than the value derived
from Monte Carlo calculations of $\alpha=0.69$ \citep{vikolamet}. Models based on
the comoving-frame radiative transfer calculations \citep{mcmfkont} give even
lower value of $\alpha=0.59$, which is close to the canonical value mentioned
above ($\alpha=0.5$). Their mass-loss rate scaling with metallicity also agrees
with empirical estimates \citep{marcozet}.

The terminal velocity, $\vinfty$, is another important wind parameter, which gives
the speed of the wind at large distances from the star. Unlike the mass-loss
rate, the wind terminal velocity only weakly depends on the metallicity and is
mostly given by the escape speed from the stellar surface multiplied by the
factor $\hat\alpha/(1-\hat\alpha)$ \citep{cak}. Therefore, the metallicity
dependence of the terminal velocity stems from the variations of the
$\hat\alpha$ parameter with metallicity. Detailed calculations do not offer any
clear predictions concerning the dependence of wind terminal velocity on the
metallicity \citep{mbcmfkont}, partly because the terminal velocity is given by
the integration of equation of motion over large interval of radii. Therefore,
wind terminal velocity might be strongly affected by uncertainties of the
radiative force connected, for example, with the influence of X-rays or clumping
\citep[e.g.,][]{lojza,irchuch,savi}.

The observational analysis shows that the wind terminal velocity is correlated not
only with the escape speed, but also with the stellar effective temperature
\citep{prina,snadtostihnou}. Moreover, the observations indicate slight increase
in the wind terminal velocity with metallicity, $v_\infty\sim Z^{0.22}$
\citep{snadtostihnou}. While the variation in the terminal speed with effective
temperature can be understood as coming from evolutionary effects \citep{kupul,mbcmfkont},
the theoretical predictions of the wind terminal velocity variations with
metallicity are ambiguous. \citet{mcmfkont} did not predict any strong
correlation between the wind terminal velocity and metallicity; whereas
\citet{savi} showed an increase in the wind terminal velocity with metallicity for
optically thick winds of helium stars and the calculations from \citet{bjorko}
offered predictions of different slopes for the variations in the dependence on luminosity. In any
case, any variations of the wind terminal velocity with metallicity are likely
weak, as exemplified in the case of stars from IC~1613, which are expected to
have even lower metallicity than stars from the Large and Small Magellanic
Clouds (LMC and SMC), while still exhibiting similar terminal velocities \citep{garmic}.

Theoretical wind parameter estimates are typically confined to metallicities
corresponding to our Galaxy and the LMC and SMC
\citep[e.g.,][]{alexvyvoj,bjorevol,mbcmfkont} corresponding to the range of
metallicities of $0.2\leq Z/Z_\odot\leq1$. This metallicity range is well covered
also by observational surveys, such as ULLYSES \citep{ulisne2} and XShootU
\citep{xshootu}. However, observational studies become possible also for stars
at even lower metallicities \citep{garmic,zalchudkov}. The theoretical wind
predictions for stars at even lower metallicities than that corresponding to the
SMC are scarce. Therefore, to enable test of the wind models for even lower
metallicities, we provide  mass-loss rate predictions for OB stars with
metallicities down to $0.01Z/Z_\odot$ based on our METUJE code.

\section{Wind models}

State-of-the-art mass-loss rate predictions in hot stars are based on global
(unified) wind models \citep{grahamz,cmfkont,sundyn}, which solve the radiative
transfer equation line by line from IR to UV regions and
self-consistently account for the interaction between the stellar wind and
photosphere. In particular, we applied our METUJE global wind code
\citep{cmfkont}, which solves wind equations from nearly hydrostatic photosphere
to supersonically expanding wind, while assuming spherical symmetry and stationarity.

The METUJE code determines radial variations of wind density, radial velocity,
and temperature by solving corresponding hydrodynamical equations together with
the radiative transfer and kinetic (statistical) equilibrium equations (usually
referred to as the NLTE modeling). This allows us to consistently predict wind
mass-loss rate from the equation of continuity and wind terminal speed from the
equation of motion. The radiative force and radiative heating and cooling terms
are derived from the solution of the radiative transfer equation, which is
solved frequency by frequency in the comoving frame \citep{mikuh}. The
calculated radiation field is further used to determine the influence of the
radiation processes on the level populations via kinetic equilibrium equations
\citep {hubenymihalas}. These equations are solved for most abundant elements
\citep[listed in][]{btvit}, which can significantly contribute either to the
radiative driving or heating. The models are calculated assuming a smooth wind
neglecting small-scale structure (clumping).

METUJE solves all equations iteratively allowing us to determine a consistent
solution of wind equations. The initial estimate of atmospheric structure is
derived from TLUSTY stellar atmosphere code \citep{ostar2003,bstar2006}. We
typically calculate the TLUSTY models for the same parameters as METUJE models; however, for subMagellanic metallicities, using models with SMC
composition was sufficient in most cases.

In this work, the models were parameterized by the stellar effective temperature, \teff, stellar
mass, $M$, radius, $R_*$, and surface chemical composition. The resulting grid of
wind parameters covers O main-sequence stars, giants, and supergiants and B
supergiants. For O stars, we selected the same grid of stellar parameters as in
\citet{mcmfkont}, which was derived using relations of \citet{okali}. The grid
for B supergiants was calculated at three different values of stellar luminosity,
$L$, for the same stellar parameters as in \citet{bcmfkont}. The selected
parameters correspond to typical values determined for B supergiants from our
Galaxy \citep{vysbeta}. The models were calculated for three different
metallicities $Z_\odot/10$, $Z_\odot/30$, and $Z_\odot/100$ derived by scaling
solar chemical composition ($Z_\odot$) of \citet{asp09} for each individual
element.

\section{Calculated wind parameters}

Predicted values of wind mass-loss rates and terminal velocities are given in
Table~\ref{bvele} for B supergiants and in Table~\ref{ohvezpar} for O stars. For
the subsequent analysis, these models were complemented with OB star wind models
at Galactic ($Z_\odot$), LMC ($Z_\odot/2$), and SMC ($Z_\odot/5$) metallicities
calculated earlier with the same code for the same stellar masses, luminosities,
and radii \citep{cmfkont,mcmfkont,bcmfkont,mbcmfkont}.

\begin{figure}
\includegraphics[width=0.5\textwidth]{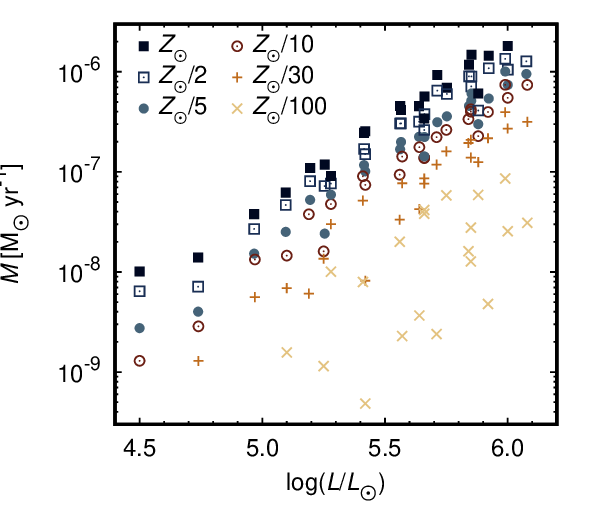}
\caption{Wind mass-loss rate as a function of the stellar luminosity from METUJE
models at different metallicities for stars with
$T_\text{eff}\geq27.5\,\text{kK}$.}
\label{dmdtlgls}
\end{figure}

\begin{figure}
\includegraphics[width=0.5\textwidth]{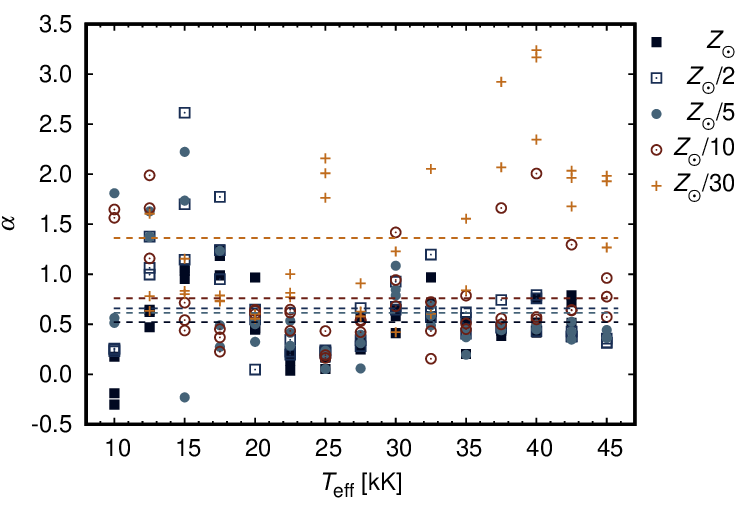}
\caption{Power of the metallicity dependence of the mass-loss rate $\dot
M\sim Z^\alpha$ (Eq.~\ref{alfa}) plotted as a function of the stellar effective
temperature for models at different metallicities. Dashed lines describe mean
relationship derived for individual metallicities.}
\label{dmdtalfa}
\end{figure}

\begin{figure}
\includegraphics[width=0.5\textwidth]{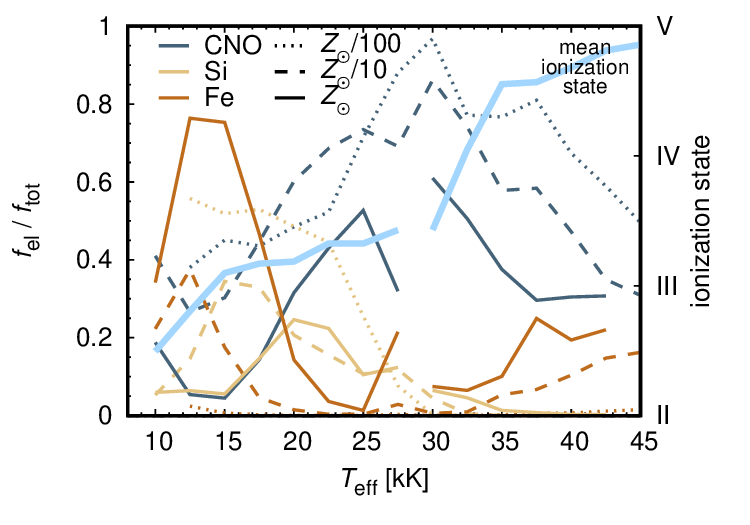}
\caption{Relative contribution of selected elements to the line radiative force
as a function of the effective temperature, plotted at the wind critical point
for O giants and B supergiants with $M=40\,M_\odot$. While line colors denote
individual elements, line styles differentiate between metallicity, with the
solid line corresponding to $Z_\odot$, dashed line to $0.1Z_\odot$, and dotted
line to $0.01Z_\odot$. The thick blue line gives the mean ionization state
driving the wind for $0.1Z_\odot$. The mean ionization state is defined as $\sum
f_iz_i/(\sum f_i)$, where $f_i$ is the contribution of ion $i$ (with charge
$z_i$) to the line radiative force. The corresponding axis is on the right. The
values of the radiative force were obtained within the Sobolev approximation.}
\label{zsil}
\end{figure}

\begin{table*}[t]
\caption{Predicted terminal velocities $v_\infty$ (in \kms) and the mass-loss
rates $\dot M$ (in \msr) for B supergiants.}
\centering
\label{bvele}
\begin{tabular}{ccccccccc}
\hline
\hline
&&&\multicolumn{2}{c}{$Z=Z_\odot/10$}&
\multicolumn{2}{c}{$Z=Z_\odot/30$}&
\multicolumn{2}{c}{$Z=Z_\odot/100$}\\
Model &$\teff$ $[\text{kK}]$ & $R_{*}$ $[{R}_{\odot}]$ &
$\vinfty$ & $\dot M$ &
$\vinfty$ & $\dot M$ &
$\vinfty$ & $\dot M$ \\
\hline
\multicolumn{9}{c}{$M=25\,{M}_{\odot}$, $\log(L/L_\odot)=5.28$,
$\Gamma=0.18$}\\
275-25 & 27.5 & 19.3 & 1640 & $4.8\times10^{-8}$ & 1270 & $3.0\times10^{-8}$ & 1100 & $1.0\times10^{-8}$\\ 
250-25 & 25.0 & 23.3 & 1130 & $4.8\times10^{-8}$ &  960 & $3.0\times10^{-8}$ & 1150 & $2.6\times10^{-9}$\\
225-25 & 22.5 & 28.8 & 1220 & $3.8\times10^{-8}$ & 1240 & $1.9\times10^{-8}$ &  980 & $7.3\times10^{-9}$\\
200-25 & 20.0 & 36.4 & 1260 & $2.7\times10^{-8}$ & 1160 & $1.3\times10^{-8}$ &  760 & $6.3\times10^{-9}$\\
175-25 & 17.5 & 47.6 & 1100 & $1.8\times10^{-8}$ &  830 & $1.1\times10^{-8}$ &  160 & $4.2\times10^{-9}$\\
150-25 & 15.0 & 64.8 &  630 & $1.6\times10^{-8}$ &  670 & $7.2\times10^{-9}$ &  410 & $1.8\times10^{-9}$\\
125-25 & 12.5 & 93.3 &   90 & $1.1\times10^{-8}$ &  110 & $3.1\times10^{-9}$ &   60 & $4.5\times10^{-10}$\\
100-25 & 10.0 & 146  &  330 & $2.5\times10^{-9}$ &\multicolumn{2}{c}{no wind}&\multicolumn{2}{c}{no wind}\\
\hline
\multicolumn{9}{c}{$M=40\,{M}_{\odot}$, $\log(L/L_\odot)=5.66$,
$\Gamma=0.27$}\\
275-40 & 27.5 & 29.9 & 1800 & $1.4\times10^{-7}$ & 1550 & $7.7\times10^{-8}$ & 1170 & $3.8\times10^{-8}$\\
250-40 & 25.0 & 36.1 & 1250 & $1.3\times10^{-7}$ &  810 & $1.1\times10^{-7}$ & 1180 & $1.3\times10^{-8}$\\
225-40 & 22.5 & 44.6 &  940 & $1.2\times10^{-7}$ & 1140 & $6.1\times10^{-8}$ & 1080 & $2.3\times10^{-8}$\\ 
200-40 & 20.0 & 56.4 & 1160 & $7.9\times10^{-8}$ & 1140 & $4.0\times10^{-8}$ &  830 & $2.0\times10^{-8}$\\
175-40 & 17.5 & 73.7 & 1210 & $4.3\times10^{-8}$ &  970 & $3.4\times10^{-8}$ &  680 & $1.4\times10^{-8}$\\
150-40 & 15.0 & 100  &  130 & $4.6\times10^{-8}$ &  180 & $2.5\times10^{-8}$ &  140 & $9.3\times10^{-9}$\\ 
125-40 & 12.5 & 145  &  110 & $6.7\times10^{-8}$ &   40 & $7.5\times10^{-9}$ &   90 & $3.5\times10^{-9}$\\
100-40 & 10.0 & 226  &  320 & $1.9\times10^{-8}$ &  260 & $3.1\times10^{-9}$ &\multicolumn{2}{c}{no wind}\\
\hline
\multicolumn{9}{c}{$M=60\,{M}_{\odot}$, $\log(L/L_\odot)=5.88$,
$\Gamma=0.30$}\\
275-60 & 27.5 & 38.5 & 1870 & $2.3\times10^{-7}$ & 1660 & $1.3\times10^{-7}$ & 1260 & $5.9\times10^{-8}$\\ 
250-60 & 25.0 & 46.5 & 1450 & $2.2\times10^{-7}$ &  920 & $1.8\times10^{-7}$ & 1380 & $1.4\times10^{-8}$\\
225-60 & 22.5 & 57.5 & 1090 & $1.8\times10^{-7}$ & 1090 & $1.1\times10^{-7}$ & 1180 & $3.3\times10^{-8}$\\
200-60 & 20.0 & 72.7 & 1270 & $1.4\times10^{-7}$ & 1270 & $7.2\times10^{-8}$ &  990 & $3.5\times10^{-8}$\\
175-60 & 17.5 & 95.0 & 1290 & $8.5\times10^{-8}$ & 1040 & $5.7\times10^{-8}$ &  760 & $2.3\times10^{-8}$\\
150-60 & 15.0 & 129  &  640 & $8.3\times10^{-8}$ &  680 & $5.1\times10^{-8}$ &  550 & $2.0\times10^{-8}$\\ 
125-60 & 12.5 & 186  &  370 & $1.2\times10^{-7}$ &   80 & $1.9\times10^{-8}$ &   90 & $7.4\times10^{-9}$\\
100-60 & 10.0 & 291  &  110 & $3.2\times10^{-8}$ &   80 & $5.8\times10^{-9}$ &\multicolumn{2}{c}{no wind}\\
\hline
\end{tabular}
\tablefoot{$\Gamma$ is the Eddington parameter for electron Thomson scattering.
For simplicity, $\Gamma$ is calculated assuming fully ionized medium.}
\end{table*}

\begin{table*}
\caption{Predicted terminal velocities $v_\infty$ (in \kms) and the mass-loss
rates $\dot M$ (in \msr) for O stars.}
\centering
\label{ohvezpar}
\begin{tabular}{ccrccccccccc}
\hline
\hline
&&&&&&\multicolumn{2}{c}{$Z=Z_\odot/10$}&
\multicolumn{2}{c}{$Z=Z_\odot/30$}&
\multicolumn{2}{c}{$Z=Z_\odot/100$}\\
Model &$\Teff$ $[\text{K}]$ & $R_{*}$ $[\text{R}_{\odot}]$ &
$M$
$[\text{M}_{\odot}]$
& $\log(L/L_\odot)$ & $\Gamma$ 
& $\vinfty$
& $\dot M$ & $\vinfty$ & $\dot M$ & $\vinfty$ &
$\dot M$\\
\hline \multicolumn{12}{c}{main sequence stars}\\
300-5 & 30000 & 6.6 & 12.9  & 4.50 & 0.06 & 1780 & $1.3\times10^{-9}$ &  990 & $2.7\times10^{-10}$& \multicolumn{2}{c}{no wind}\\
325-5 & 32500 & 7.4 & 16.4  & 4.74 & 0.08 & 1890 & $2.9\times10^{-9}$ & 1150 & $1.3\times10^{-9}$ & \multicolumn{2}{c}{no wind}\\
350-5 & 35000 & 8.3 & 20.9  & 4.97 & 0.11 & 1250 & $1.3\times10^{-8}$ &  930 & $5.6\times10^{-9}$ & \multicolumn{2}{c}{no wind}\\
375-5 & 37500 & 9.4 & 26.8  & 5.19 & 0.14 & 1790 & $3.8\times10^{-8}$ & 1260 & $6.1\times10^{-9}$ &  420 & $3.4\times10^{-11}$\\
400-5 & 40000 & 10.7 & 34.6 & 5.42 & 0.18 & 1900 & $7.4\times10^{-8}$ & 1510 & $8.2\times10^{-9}$ &  820 & $4.8\times10^{-10}$\\
425-5 & 42500 & 12.2 & 45.0 & 5.64 & 0.23 & 1850 & $1.8\times10^{-7}$ & 1260 & $4.3\times10^{-8}$ & 1240 & $3.7\times10^{-9}$ \\
450-5 & 45000 & 13.9 & 58.6 & 5.85 & 0.29 & 1830 & $4.0\times10^{-7}$ & 1300 & $1.4\times10^{-7}$ & 1490 & $1.3\times10^{-8}$ \\
\hline \multicolumn{12}{c}{giants}\\
300-3 & 30000 & 13.1 & 19.3 & 5.10 & 0.15 & 1850 & $1.5\times10^{-8}$ & 1360 & $6.9\times10^{-9}$ &  730 & $1.6\times10^{-9}$ \\
325-3 & 32500 & 13.4 & 22.8 & 5.25 & 0.19 & 2180 & $1.6\times10^{-8}$ & 1280 & $1.4\times10^{-8}$ &  510 & $1.1\times10^{-9}$ \\
350-3 & 35000 & 13.9 & 27.2 & 5.41 & 0.23 & 1380 & $9.0\times10^{-8}$ & 1050 & $5.2\times10^{-8}$ &  680 & $7.9\times10^{-9}$ \\
375-3 & 37500 & 14.4 & 32.5 & 5.57 & 0.27 & 1710 & $1.4\times10^{-7}$ & 1320 & $7.7\times10^{-8}$ & 1010 & $2.3\times10^{-9}$ \\
400-3 & 40000 & 15.0 & 39.2 & 5.71 & 0.31 & 1860 & $2.2\times10^{-7}$ & 1360 & $1.2\times10^{-7}$ & 1360 & $2.4\times10^{-9}$ \\
425-3 & 42500 & 15.6 & 47.4 & 5.85 & 0.36 & 1780 & $4.2\times10^{-7}$ & 1320 & $2.1\times10^{-7}$ &  990 & $2.8\times10^{-8}$ \\
450-3 & 45000 & 16.3 & 57.7 & 5.99 & 0.40 & 1730 & $7.4\times10^{-7}$ & 1310 & $3.9\times10^{-7}$ &  870 & $8.6\times10^{-8}$ \\
\hline \multicolumn{12}{c}{supergiants}\\
300-1 & 30000 & 22.4 & 28.8 & 5.56 & 0.30 & 1320 & $9.4\times10^{-8}$ & 1430 & $3.3\times10^{-8}$ &  900 & $2.0\times10^{-8}$ \\
325-1 & 32500 & 21.4 & 34.0 & 5.66 & 0.32 &  940 & $1.4\times10^{-7}$ & 1020 & $8.6\times10^{-8}$ &  690 & $4.2\times10^{-8}$ \\
350-1 & 35000 & 20.5 & 40.4 & 5.75 & 0.33 & 1540 & $2.6\times10^{-7}$ & 1250 & $1.6\times10^{-7}$ &  920 & $5.8\times10^{-8}$ \\
375-1 & 37500 & 19.8 & 48.3 & 5.84 & 0.34 & 1680 & $3.4\times10^{-7}$ & 1400 & $1.9\times10^{-7}$ & 1050 & $1.6\times10^{-8}$ \\
400-1 & 40000 & 19.1 & 58.1 & 5.92 & 0.34 & 1910 & $4.0\times10^{-7}$ & 1480 & $2.2\times10^{-7}$ & 1630 & $4.8\times10^{-9}$ \\
425-1 & 42500 & 18.5 & 70.3 & 6.00 & 0.34 & 1980 & $5.5\times10^{-7}$ & 1440 & $2.7\times10^{-7}$ & 1340 & $2.6\times10^{-8}$ \\
450-1 & 45000 & 18.0 & 85.4 & 6.08 & 0.33 & 1950 & $7.4\times10^{-7}$ & 1460 & $3.2\times10^{-7}$ & 1660 & $3.1\times10^{-8}$ \\
\hline
\end{tabular}
\end{table*}

In general, the mass-loss rate increases with increasing luminosity and
metallicity; in addition, it depends on the effective temperature. With
increasing luminosity, the radiative force and, consequently, mass-loss rate
also increase (Fig.~\ref{dmdtlgls}).

The metallicity dependence of the wind mass-loss rate stems from increase in the
line opacity with increasing metallicity, which increases the line radiative
force. The decrease in the wind mass-loss rate with decreasing metallicity is
already visible from the downward shift of luminosity variations plotted for
different metallicities in Fig.~\ref{dmdtlgls}. Moreover, this plot also shows
that while the luminosity dependence of the mass-loss
rate is monotonic  for higher metallicities, for subMagellanic metallicities ($Z\lesssim0.1Z_\odot$), the
luminosity relationship shows scatter, which is a particularly considerable one for $Z =0.01Z_\odot$. The metallicity dependence of the mass-loss rate is further demonstrated in
Fig.~\ref{dmdtalfa}, where we plot the steepness of the metallicity dependence of
the mass-loss rate, $\dot M\sim Z^\alpha$, 
\begin{equation}
\label{alfa}
\alpha=\frac{\partial\log\dmdt}{\partial\log Z}
\end{equation}
as a function of the effective temperature. The parameter $\alpha$ is evaluated
for stellar models in Tables \ref{bvele} and \ref{ohvezpar}. The plot
in Fig.~\ref{dmdtalfa} approximates the derivative using backward differences
calculated for metallicities from the grid. For example, the $\alpha$ value at
the solar metallicity is approximated as $\alpha(Z_\odot)=
\log(\dmdt(Z_\odot)/\dmdt(0.5Z_\odot))/\log(Z_\odot/0.5Z_\odot)$. Consequently,
there is no value for $\alpha(0.01 Z_\odot)$. From the plot, it follows that
while at the metallicities corresponding to our Galaxy and the Magellanic Clouds
the metallicity dependence of the mass-loss rate can be approximated by a single
relationship $\dot M\sim Z^{0.60}$, for lower metallicities
($Z\lesssim0.1Z_\odot$) the relationship significantly steepens, even reaching as high
as $\dot M\sim Z^ {1.4}$ for $Z_\odot/30$.

The variations in the power $\alpha$ with the temperature are not monotonic, but they do show a maximum around $T_\text{eff}\approx 15$\,kK and a significant scatter for
subMagellanic metallicities ($Z\lesssim0.1Z_\odot$) in Fig.~\ref{dmdtalfa}. These
variations can be understood using Fig.~\ref{zsil}, where we plot
contribution of individual elements to the radiative force. The radiative
driving in the O star domain is dominated by CNO elements, while iron dominates
in B supergiants at solar metallicity \citep{vikolabis}, which is replaced by
silicon at subMagellanic abundances. These variations stem from dependence of
flux distribution and ionization degree on the effective temperature. In the O
star domain, a significant fraction of line driving originates in the spectral
region of the Lyman continuum \citep{abpar}, where ions such as \ion{N}{iv},
\ion{O}{v}, and \ion{Ne}{iv} have many strong lines driving the wind. As the
temperature decreases, the maximum of the flux distribution shifts to the Balmer
continuum, and ionization balance of CNO elements becomes dominated by
\ion{C}{iii}, \ion{N}{iii}, and \ion{O}{ii}. However, these ions also have the
strongest resonance lines in the Lyman continuum and this mismatch between the
wavelength of the maximum flux and the position of the strongest lines leads to
a decrease in the contribution of CNO elements to the radiative
force.\footnote{An exception is the \ion{C}{iii} 977\,\AA\ resonance line, which
is able to significantly contribute to the line acceleration even in the B
supergiant domain.} Instead, a significant contribution of \ion{Fe}{iii} appears
at solar metallicity, which has a large number of strong lines at longer
wavelengths. However, the iron lines become optically thin at subMagellanic
metallicities and iron contribution is replaced by \ion{Si}{iv}, whose
1394\,\AA\ and 1403\,\AA\ resonance doublet together with \ion{C}{iii} 977\,\AA\
line might provide more than two thirds of the line acceleration at the lowest
metallicities in the region of the critical point.

\begin{figure}
\includegraphics[width=0.5\textwidth]{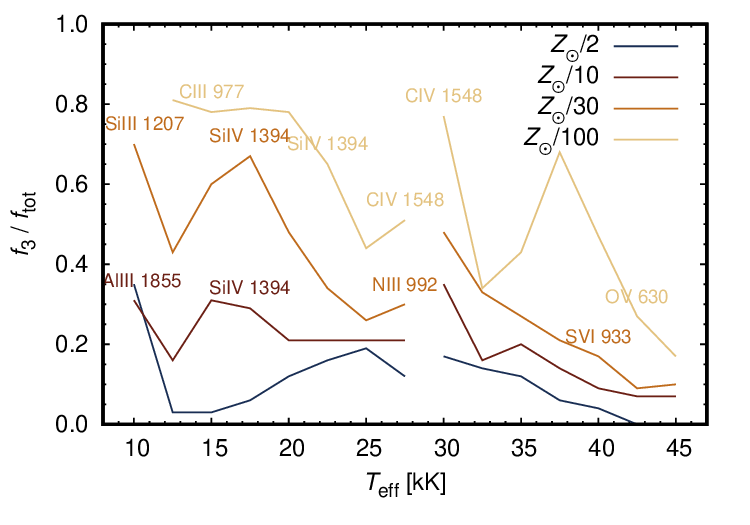}
\caption{Relative contribution of three lines that drive the wind most strongly
to the radiative force. Plotted as a function of the stellar effective
temperature for different metallicities at the critical point using the Sobolev
approximation for O giants and $40\,M_\odot$ B supergiants.
We labeled selected strongest lines with wavelength in \AA.}
\label{trisil}
\end{figure}

The contribution of individual lines to the radiative force is further illustrated
in Fig.~\ref{trisil}, which gives the cumulative fraction of the radiative force
coming from the three strongest lines. For higher metallicities, the
contribution of the three strongest lines to the radiative force is typically
below 10\%. This implies that the wind is driven by a large number of lines at
high metallicities. However, with decreasing metallicity, many of the strong lines
become optically thin and their contribution to the radiative force becomes
quenched. As a result, the wind is driven just by a few lines at the lowest
metallicities.

The low number of lines that significantly drive the wind at the lowest
metallicities leads to a strong variation in the mass-loss rate with stellar
parameters. This can be seen as a high dispersion in the plots of the dependence
of the mass-loss rate on luminosity in Fig.~\ref{dmdtlgls} and the power of
metallicity dependence (Fig.~\ref{dmdtalfa}) for $Z\leq0.1Z_\odot$. The decrease
in the mass-loss rate with metallicity is not only caused directly by a
decreased metallicity, but also indirectly by a reduced wind density. Lower wind
density leads to ionization shift toward higher ions, which drive the wind less
efficiently, leading to decrease in the mass-loss rate. For instance, the wind
of a 400-1 star is driven to a large fraction by such ions as \ion{O}{iv} and
\ion{Ne}{iv}. With decreasing metallicity, the line force decreases, leading to
the reduction of the wind density. The decrease in the density leads to higher
ionization, diminishes the ionization fraction of \ion{O}{iv} and \ion{Ne}{iv}
and further reduces the mass-loss rate. As a result, the metallicity power
$\alpha$ becomes very high around $T_\text{eff}\approx40\,\text{kK}$. Below and
above this temperature, the changes due to ionization are not so high; therefore,
$\alpha$ is lower.

A high value of the metallicity power $\alpha$ at
$T_\text{eff}\approx15\,\text{kK}$ in Fig.~\ref{dmdtalfa} is caused by yet
another effect. Our models feature so-called bistability bump at solar
metallicity in this temperature domain \citep{mnichov8}, which is caused by
increase in the line force due to iron recombination from \ion{Fe}{iv} to
\ion{Fe}{iii} \citep{vikolabis,vinbisja,bcmfkont}. With decreasing metallicity,
the iron lines become optically thin, their contribution to the radiative force
decreases (Fig.~\ref{zsil}) and the bistability jump vanishes. As a result,
$\alpha$ is higher than one for $Z\gtrsim0.2Z_\odot$ around
$T_\text{eff}\approx15\,\text{kK}$.

\begin{figure}
\includegraphics[width=0.5\textwidth]{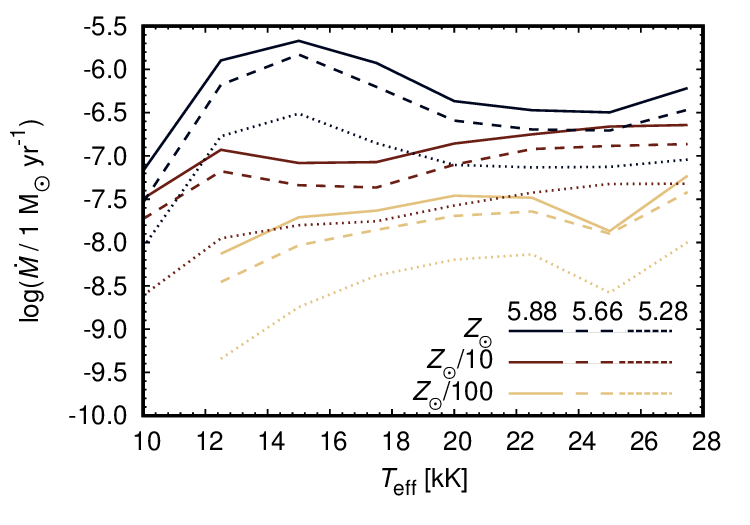}
\caption{Mass-loss rate variations with the temperature for B supergiants. The line
color denotes metallicity, while the dash type corresponds to different stellar
luminosities ($\log(L/L_\odot)=5.28,\,5.66,$ and 5.88).}
\label{zdmdttep}
\end{figure}

The metallicity variations of the bistability jump are apparent also in
Fig.~\ref{zdmdttep} showing the mass-loss rate as a function of the effective
temperature for different stellar luminosities. At solar metallicity, a clear
maximum appears at about 15\,kK, which weakens for lower metallicities and
shifts toward lower temperatures \citep[as follows from their Eqs. (14) and
(15)]{vikolamet}. The shift appears due a stronger ionization in winds with
lower densities. The extended width of the bistability region results from
gradual dependence of recombination of \ion{Fe}{iv} on temperature. The latter
ion dominates at high effective temperatures, but at 20\,kK the first part of
the wind dominated by \ion{Fe}{iii} appears. \ion{Fe}{iii} starts to dominate
the whole wind at $T_\text{eff}\approx15\,\text{kK}$ or even cooler at low
metallicities. \citet{petcmfgen} found a narrow bistability jump and at a higher
effective temperature of 21\,kK. This is likely the result of their adopted
terminal velocity variation, which is based on the empirical results of \citet{lsl},
featuring a jump. \citet{vysbeta} found smooth variations of the ratio of the
wind terminal velocities and the escape speed, which likely smooths the jump in
mass-loss rates. \citet{vinbisja} also found smooth bistability jump and
increasing wind mass-loss rate even below 20\,kK from the consistent solution of
the momentum equation. Calculations of \citet{bjorevol} do not predict any
bistability jump from their models. The reason for this absence is not fully
clear.

\begin{table}[t]
\caption{Parameters of the fit of the mass-loss rate in
Eq.~\eqref{dmdtobsloppy}.}
\label{sloppy}
\begin{tabular}{*{7}{c}}
\hline
\hline
$\tilde a$ & $\tilde b$ & $\tilde c$ & $\tilde d$ & $\tilde e$ &
$\tilde T_0$ [kK] & $\Delta\tilde T$ [kK] \\
\hline
 $-7.72$ & 1.49 & 0.713 & 1.29 & 1.10 & 14.4 & 2.53 \\
\hline
\end{tabular}
\centering
\end{table}

As a result of strong variations of the mass-loss rate with stellar parameters
(mainly temperature), we were unable to derive a precise fit of the predicted 
mass-loss rates. Nonetheless, the mass-loss rates can be roughly fitted via a
rather sloppy fit as
\begin{multline}
\label{dmdtobsloppy}
\log\zav{\frac{\dot M}{1\, \msr }}= \tilde a +\tilde b
\log\zav{\frac{L}{10^6L_\odot}} +\tilde c\log\zav{\frac{Z}{Z_\odot}}+
\tilde d\log\zav{\frac{T_\text{eff}}{10^3\,\text{K}}}\\+
\tilde e\zav{\frac{Z}{Z_\odot}}
\exp\hzav{-\frac{(T_\text{eff}-\tilde T_0)^2}{\Delta\tilde T^2}}.
\end{multline}
Throughout our paper, $\log$ denotes decadic logarithm and $\exp$ the natural
exponential function. Parameters of the fit are given in Table~\ref{sloppy}. The
fit does not represent the predicted mass-loss rates particularly well with root
mean square (rms) of the residual of about $ 0.28 \,$dex and a maximum deviation for
individual stars as high as a factor of 3. 

\begin{table}[t]
\caption{Parameters of the fit of the mass-loss rate in Eq.~\eqref{dmdtob}.}
\label{fit}
\begin{tabular}{*{4}{c}}
\hline
\hline
$a$ & $b$ & $T_0$ [kK] & $\Delta T$ [kK] \\
\hline
$-7$ & $1.619$ & 10 & 35\\
\hline
\end{tabular}
\centering
\bigskip

\begin{tabular}{rrrrr}
\hline
\hline
$l_{nm}$ & $m=0$ & $m=1$ & $m=2$ & $m=3$\\
\hline
$n=1$ & $-31.333$  & $-46.3518$ & $-19.0704$ & 0.372222\\
$n=2$ & $119.62$   & $170.353$  & $70.4817$  & 0.119759\\
$n=3$ & $-201.713$ & $-284.575$ & $-115.793$ & 0\\
$n=4$ & $208.466$  & $292.063$  & $117.131$  & 0\\
$n=5$ & $-140.704$ & $-195.337$ & $-77.1403$ & 0\\
$n=6$ & $58.7226$  & $80.3569$  & $31.2397$  & 0\\
$n=7$ & $-11.7687$ & $-15.3933$ & $-5.729$   & 0\\
\hline
\end{tabular}
\centering
\end{table}

A fussier fit can be derived using Legendre polynomials%
\footnote{The code is available at
\href{https://zenodo.org/records/15965163}
{https://zenodo.org/records/15965163}.}
\begin{multline}
\label{dmdtob}
\log\zav{\frac{\dot M}{1\, \msr }}= a + b \log\zav{\frac{L}{10^6L_\odot}} \\
+\sum_{n=1}^7 \zav{l_{n0}+l_{n1}\log\frac{Z}{Z_\odot}+
l_{n2}\log^2\frac{Z}{Z_\odot}+l_{n3}\log^3\frac{Z}{Z_\odot}}\\\times
P_n\zav{\frac{T_\text{eff}-T_0}{\Delta T}},
\end{multline}
where $P_n(x)$ are Legendre polynomials of a degree, $n$, $P_n(x) =
\frac{\de^n}{\de x^n} (x^2 -1)^n/(2^n n!)$. The individual parameters are given
in Table~\ref{fit}. This  this fit does not represent the predicted mass-loss
rates very well, with a rms of the residual of about $ 0.17
\,$dex, while the deviations for some stars could be as high as a factor of 2.
The differences between the mass-loss rates derived from the models and from the
fit Eq.~\eqref{dmdtob} increase toward lower metallicities. The fits in Eqs.
\eqref{dmdtobsloppy} and \eqref{dmdtob} were derived for abundances of
$Z=0.01-1\,Z_\odot$ and effective temperatures of $10-45\,\text{kK; thus,}$  their
usage outside this range of parameters can lead to erroneous results.

\begin{figure}
\includegraphics[width=0.5\textwidth]{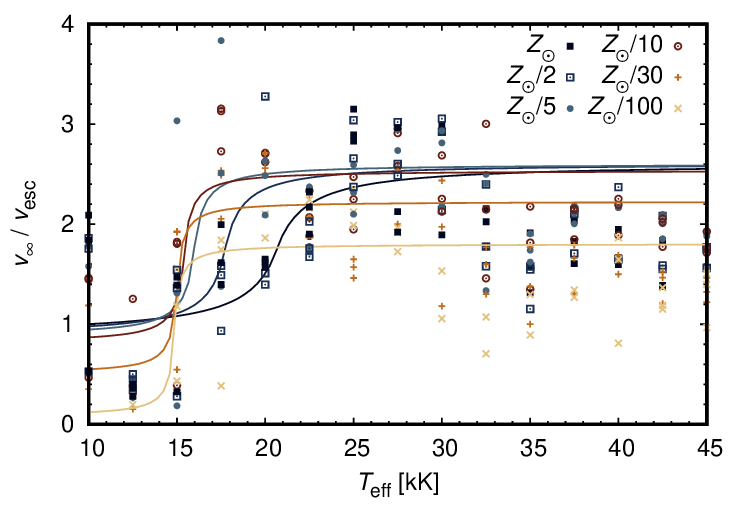}
\caption{Predicted ratio of the wind terminal velocity, $\vinfty$, and escape
speed, $\vesc$, as a function of the effective temperature. Solid lines colored
according to the metallicity give fit to theoretical predictions
(see Eq.~\eqref{vfit1}).}
\label{vnekvuniz}
\end{figure}

\begin{table}[t]
\caption{Parameters of the terminal velocity fit 
in Eqs.~\eqref{vfit1}--\eqref{vfit3}.}
\label{vfittab}
\centering
\begin{tabular}{ccccccccc}
\hline\hline
$v_{0,+}$ & $v_{0,-}$ & $\delta v$ & $\beta$ & $T_v$ & $T_{v,z}$ &
$\delta T$ & $\delta T_z$ \\
&&&&\multicolumn{4}{c}{[kK]}\\
\hline
2.61 & 0.88 & 0.49 & 8.3 & 14.8 & 5.8 & 0.2 & 1.4 \\
\hline
\end{tabular}
\end{table}

The wind terminal velocity depends mostly on the escape speed and on stellar
parameters such as stellar temperature and metallicity \citep{cak,pusle}.
Figure~\ref{vnekvuniz} demonstrates that this relationship is not smooth, but
shows a considerable scatter. On average, the terminal velocity is nearly
independent on metallicity for OB stars above the bistability jump
($T_\text{eff}\geq20\,\text{kK}$) with $Z\geq0.2Z_\odot$, while it scales
roughly as $v_\infty\sim Z^{0.18}$ for lower metallicities. On average, the
ratio of the terminal velocity to the escape speed can be approximated as
\begin{align}
\label{vfit1}
\frac{v_\infty}{v_\text{esc}}&=
\frac{1}{2}\hzav{\zav{v_+-v_-}\frac{t}{1+|t|}+v_++v_-},\\
v_{\pm}&=v_{0,\pm}+\delta v\,10^{-\beta Z/Z_\odot}\log(Z/Z_\odot),\\
\label{vfit3} t&=\frac{T_\text{eff}-T_v - T_{v,z}\frac{Z}{Z_\odot}}
{\delta T+\delta T_z\frac{Z}{Z_\odot}},
\end{align}
where the fit parameters are given in Table~\ref{vfittab}. The fit was derived
for stars with $ T_\text{eff}\leq30\,\text{kK}$. Regarding the upper limit, the
wind terminal velocity of stars with $T_\text{eff}\geq30\,\text{kK}$ becomes
affected by clumping \citep{irchuch}; consequently, we disregarded this region
for the fit.

\begin{figure}
\includegraphics[width=0.5\textwidth]{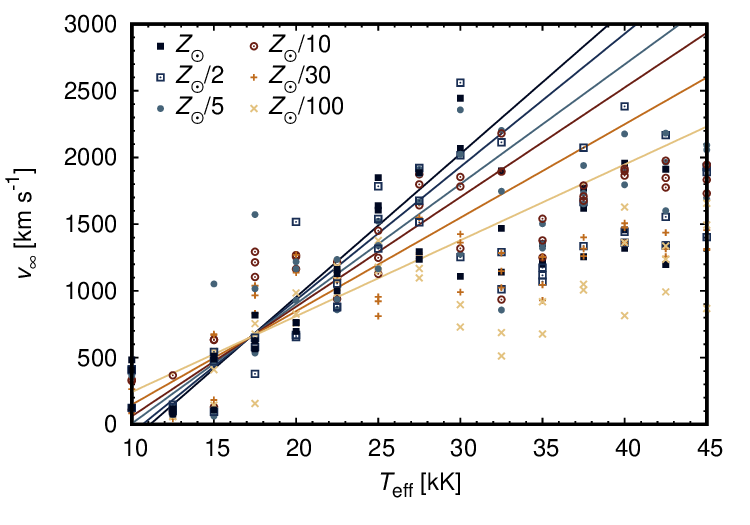}
\caption{Predicted wind terminal velocity as a function of the effective
temperature for different metallicities. The linear fits are colored according to metallicity and derived from Eq.~\eqref{vnektepzfit}.}
\label{vnektepz}
\end{figure}

\begin{table}[t]
\caption{Fit parameters  for the wind terminal velocity as a function of
the effective temperature, from Eq.~\eqref{vnektepzfit}.}
\label{vnektepztab}
\centering
\begin{tabular}{cccc}
\hline
\hline
$v_a$ & $v_{az}$ & $v_b$ & $v_{bz}$\\
\hline
107 & 25 & $-1190$ & $-430$\\
\hline
\end{tabular}
\end{table}

From the observational analysis, it follows that the wind terminal velocity scales
with the effective temperature \citep{prina,snadtostihnou}. We plot this
relation derived from our models in Fig.~\ref{vnektepz}. The theoretical
predictions can be fitted as
\begin{equation}
\label{vnektepzfit}
\zav{\frac{v_\infty}{\kms}}=\hzav{v_a+v_{az}\log\zav{\frac{Z}{Z_\odot}}}
\zav{\frac{T_\text{eff}}{10^3\,\text{K}}}+v_b+v_{bz}\log\frac{Z}{Z_\odot}.
\end{equation}
The parameters of the fit are given in Table~\ref{vnektepztab}. These parameters
were derived for $12.5\,\text{kK}\leq T_\text{eff}\leq30\,\text{kK}$.

\section{Comparison with observational estimates}

There are several diagnostics that can be used to determine the wind mass-loss
rates from observations, including near-infrared (NIR) line spectroscopy
\citep{najaro}, optical, and ultraviolet (UV) spectroscopy \citep{bouhil,clres2}, X-ray line
profiles \citep{cohcar}, and  wind bow shocks \citep{kobul}. However, at low
metallicities, only results based on optical and UV analysis are typically
available. To obtain a homogeneous observational sample, we focus only on the
mass-loss rates estimated from optical and UV line spectroscopy that accounts
for wind clumping. These values were obtained typically within the ULLYSES
\citep{ulisne2} and XShootU \citep{xshootu} projects in the case of the LMC and
SMC.

\begin{figure}
\includegraphics[width=0.5\textwidth]{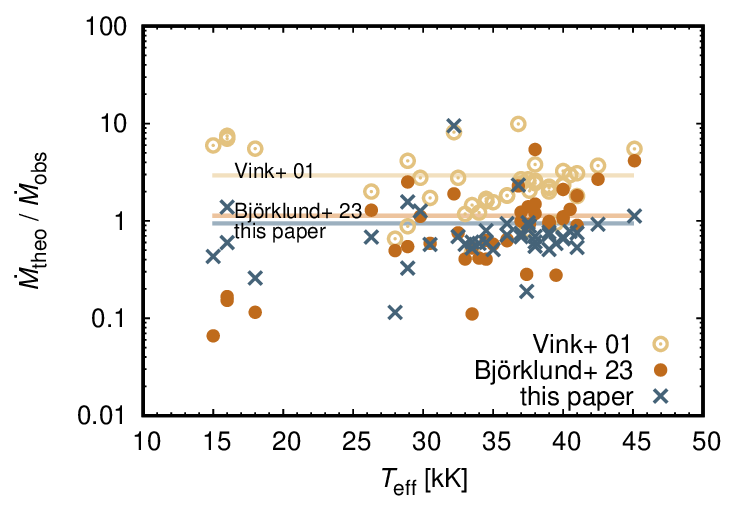}
\caption{Ratio of the predicted mass-loss rates and their observational
estimates plotted as a function of the stellar effective temperature for stars
from our Galaxy. The theoretical mass-loss rates from \citet{vikolamet},
\citet{bjorevol}, and this paper are compared with observational estimates of
\citet{najaro}, \citet{bouhil}, \citet{clres2}, \citet [supergiants]{moir},
\citet{preduli}, and \citet{berper}. The horizontal lines give mean values for
individual theoretical predictions.}
\label{dmdttepporg}
\end{figure}

Figure~\ref{dmdttepporg} compares the observational estimates with theoretical
predictions calculated for parameters of individual stars from Galaxy. On
average, the mean ratio of the theoretical values and observational
determination is 0.94, which just slightly underestimates the observations, but
the median is about 0.7. The predictions of \citet{bjorevol} give comparable
ratio of 1.13, but the mass-loss rates are underestimated by a factor of about
10 in the B star domain \citep{berper}. The predictions of \citet{vikolamet}
overestimate the mass-loss rates on average by a factor of about 2.9. The ratio
of the theoretical to observational estimates is nearly the same reported in the observational analysis of \citet{bouhil}, which assumes optically thin clumps,
as well as in \citet{clres2} and \citet{preduli}, where optically thick clumps were allowed.
This indicates that the current approach to modeling of optically thick clumps
does not introduce a systematic shift in estimated mass-loss rates
\citep{preduli,brandxii}.

\begin{figure}
\includegraphics[width=0.5\textwidth]{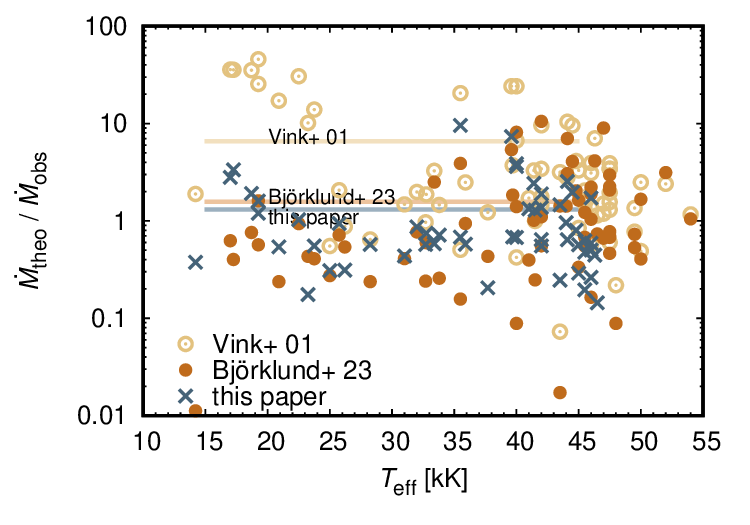}
\caption{Ratio of the predicted mass-loss rates and their observational
estimates plotted as a function of the effective temperature for stars from the
LMC. The theoretical mass-loss rates from \citet{vikolamet}, \citet{bjorevol},
and this paper are compared with observational estimates of \citet{hezkysedi},
\citet{hawvmm}, and \citet{kdejist}. The horizontal lines give mean values for
individual theoretical predictions.}
\label{dmdttepporl}
\end{figure}
Figure~\ref{dmdttepporl} gives a similar plot for the LMC. The mean
ratio of the theoretical and observational mass-loss rates is 1.3 (for stars
with $T_\text{eff}<47\,$kK). The ratio is about 1.6 for mass-loss rates of
\citet{bjorevol} and about 7 for mass-loss rates of \citet{vikolamet}.

\begin{figure}
\includegraphics[width=0.5\textwidth]{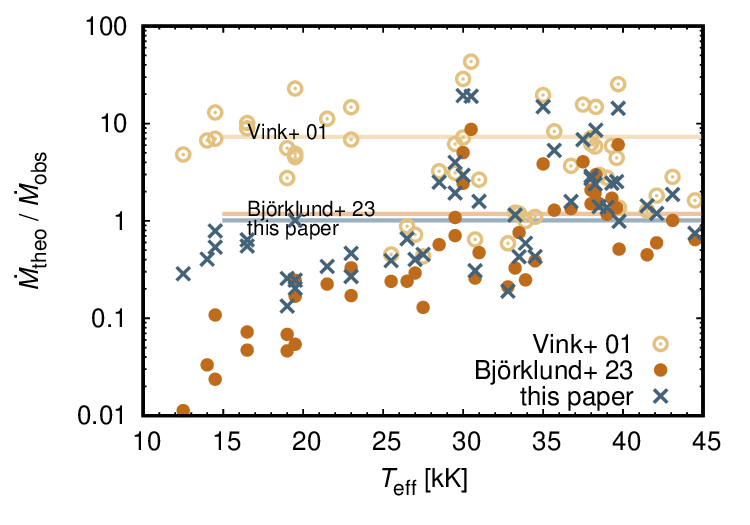}
\caption{Ratio of the predicted mass-loss rates and their observational
estimates plotted as a function of the effective temperature for stars from the
SMC. The theoretical mass-loss rates from \citet{vikolamet}, \citet{bjorevol},
and this paper are compared with observational estimates of
\citet{bousmc,bousmc2}, \citet{backss}, and \citet{berper24}.
The horizontal lines give mean values for
individual theoretical predictions.}
\label{dmdtteppors}
\end{figure}

For the SMC our mass-loss rates overestimate the observational results by a
factor of 2.8, while the rates of \citet{bjorevol} perform better and give a
mean ratio of about 1.2. However, these rates underestimate the mass-loss rates
of B supergiants by an order of magnitude \citep[see][]{berper24}, while our
mass-loss rates offer a reasonable fit. The rates of \citet{vikolamet}
overestimate the observational values by a factor of about 7. Moreover, winds of
stars with low mass-loss rates may be affected by weak-wind problem
(Sect.~\ref{wwp}). By disregarding stars with observational mass-loss rates lower
than $3\times10^{-8}\,\msr$, we obtain a much more favorable ratio of the
theoretical and observational mass-loss rates of about 1.0
(Fig.~\ref{dmdtteppors}). On this basis, we can conclude that the poor agreement between
our predictions and observational estimates for the SMC only applies to stars
with a very low mass-loss rate and is likely caused by the inefficient shock
cooling (Sect.~\ref{wwp}).

\begin{figure}
\includegraphics[width=0.5\textwidth]{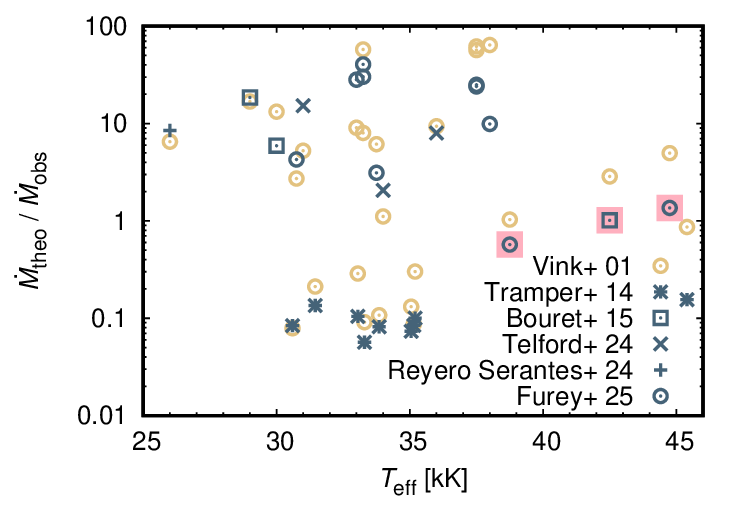}
\caption{Ratio of the predicted mass-loss rates and their observational
estimates plotted as a function of the effective temperature for stars from
galaxies with subMagellanic metallicities. Pink box denotes the only three stars
with clumping-corrected mass-loss rate $\dot M>10^{-7}\,\msr$.}
\label{dmdttepportvx}
\end{figure}

\begin{table}[t]
\caption{Mean ratio of the theoretical and observational mass-loss rates and the
rms of residuals $\chi$ defined by Eq.~\eqref{peknarovnice}.}
\label{peknatabulka}
\centering
\begin{tabular}{lcccccc}
\hline
Prediction & \multicolumn{2}{c}{Galaxy} & \multicolumn{2}{c}{LMC} &
\multicolumn{2}{c}{SMC}\\
& Ratio & $\chi$& Ratio & $\chi$& Ratio & $\chi$\\
\hline
\citeauthor{vikolamet} & 2.94 & 2.1 & 6.57 & 10  & 7.30 & 8.5\\
\citeauthor{bjorevol}  & 1.13 & 1.1 & 1.57 & 2.1 & 1.18 & 1.7\\
Current paper          & 0.94 & 1.4 & 1.31 & 1.7 & 1.01 & 0.8\\
\hline
\end{tabular}
\end{table}

These results are summarized in Table~\ref{peknatabulka}, which gives mean ratio
of theoretical mass-loss rate predictions and observational determinations
$r=\langle\dot M_\text{theo}/\dot M_\text{obs}\rangle$. In addition, it includes
the rms of residuals of this ratio, 
\begin{equation}
\label{peknarovnice}
\chi=\sqrt{\langle(\dot M_\text{theo}/\dot M_\text{obs}-r)^2\rangle}.
\end{equation}

Mass-loss rate estimates for galaxies with metallicities lower than
corresponding to the SMC are rare. \citet{desitka} derived mass-loss rates of
nine O-type stars in low-metallicity galaxies IC 1613, WLM, and NGC 3109 from
optical spectroscopy neglecting clumping. Three of these stars were subsequently
reanalyzed by \citet{nemracna} using UV spectroscopy. \citet{zalchudkov}
estimated mass-loss rates in three O stars in the nearby dwarf galaxies Leo P,
Sextans A, and WLM, with abundances of $0.03-0.14\,Z_\odot$. \citet{reys} derived
the mass-loss rate estimate of the X-1 donor star in Holmberg II dwarf galaxy
with a metallicity of $0.07\,Z_\odot$ \citep{zeelva}. \citet{ircan} derived wind
properties of massive stars from five galaxies with subMagellanic metallicities
down to $Z=0.03Z_\odot$. All these predictions are plotted in
Fig.~\ref{dmdttepportvx} in comparison to the theoretical estimates. Apparently,
the figure shows a relatively large scatter. As a result of weakness of these
winds, a part of this scatter is caused by the weak-wind problem
(Sect.~\ref{wwp}). On the other hand, the prediction for the only three stars
with the mass-loss rate higher than $10^{-7}\,\msr$ and determined accounting
for clumping (pink squares in Fig.~\ref{dmdttepportvx})  agree well with
observations.

\begin{figure}
\includegraphics[width=0.5\textwidth]{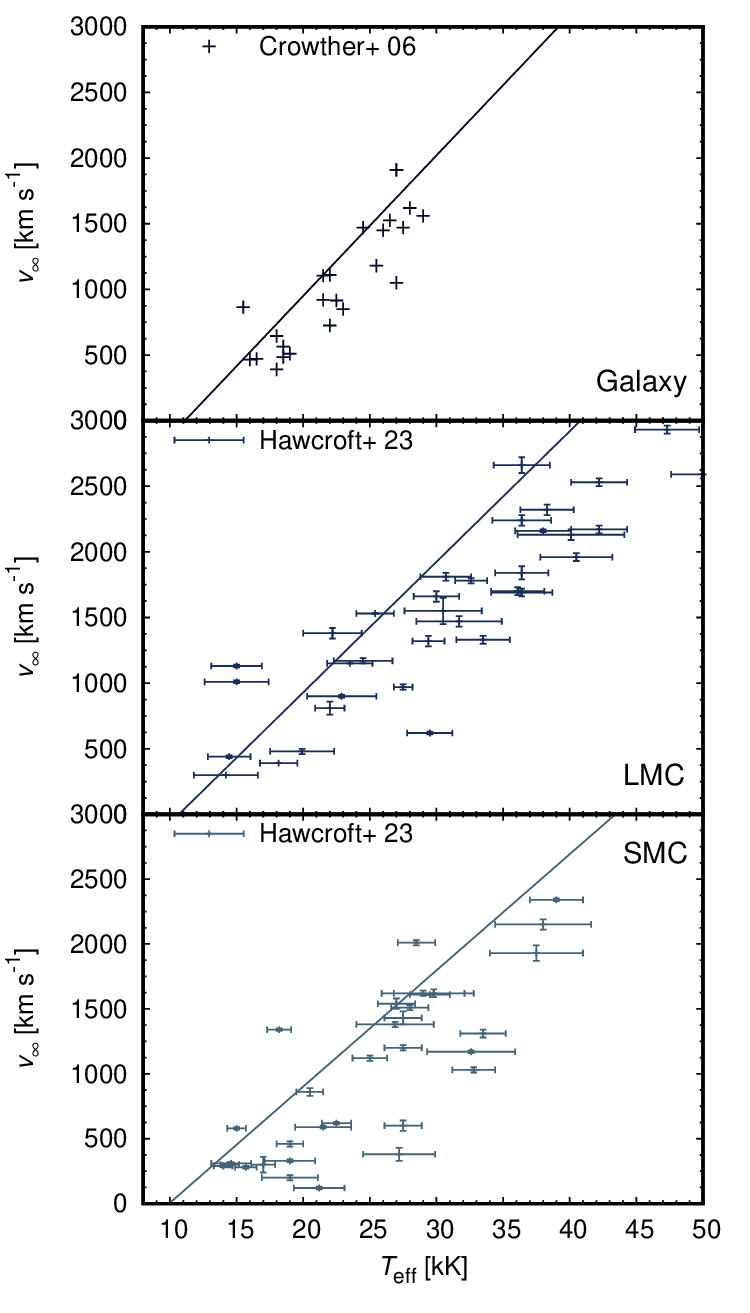}
\caption{Observational values of the wind terminal velocities
\citep{vysbeta,snadtostihnou} compared with theoretical predictions
Eq.~\eqref{vnektepzfit} for stars from the Galaxy, LMC, and SMC.}
\label{vnektep}
\end{figure}

\begin{figure}
\includegraphics[width=0.5\textwidth]{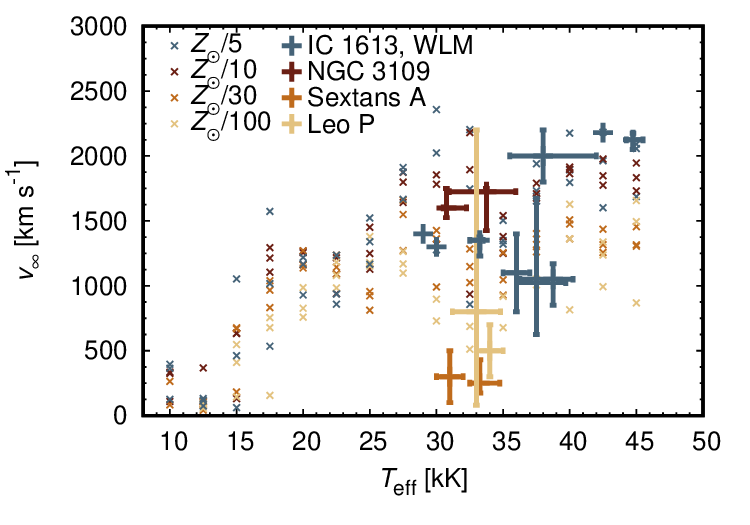}
\caption{Observational values of the wind terminal velocities
\citep{nemracna,zalchudkov,ircan} compared with theoretical values for stars
with subMagellanic metallicities.}
\label{vnekteptvx}
\end{figure}

The linear fit of wind terminal velocities Eq.~\eqref{vnektepzfit} obtained for
stars with $T_\text{eff}\leq30\,$kK reasonably agrees with observational
determinations not only in the domain of B supergiants, but also for luminous O
stars (Fig.~\ref{vnektep}). A similar comparison for galaxies with subMagellanic
metallicities shows lower than predicted terminal velocities in some stars
(Fig.~\ref{vnekteptvx}). Overall, however, the agreement is reasonable.

\section{Discussion}

\subsection{Wind velocity}

The wind radial velocity profile is often expressed in terms of the so-called
beta velocity law. The typical observational value of the $\beta$ parameter,
which determines the steepness of the radial velocity profile, is about $0.7-1$
\citep{hezkysedi,preduli}. This  agrees well with results of our simulations,
which typically give $\beta=0.6-1$. The agreement shows that theoretical models
are also able to reliably predict the radial variations of wind velocity.
However, in some cases the observational analysis gives higher values of
$\beta=1-3$ \citep{vysbeta,berper24}. Such high values of $\beta$ are typically
not found in theoretical models \citep [however, see \citealt{berperbet}]
{ppk,divnetomaj,alexcmfgen}. This discrepancy indicates that either the line
profile formation is not properly described by available spectrum-synthesis
codes \citep{pethalfa} or some physics is missing in current numerical
simulations.

Our wind models are able to reproduce the wind terminal velocities for stars
with $T_\text{eff}\leq30\,$kK (Fig.~\ref{vnektep}). For hotter stars, the
predicted terminal velocities are significantly lower than the values determined
from observations. While the Sobolev-based models with \citet{abpar} line-force
multipliers tend to overpredict the terminal velocities \citep[e.g.,][]{lsl}, in
our models, the problem of overly low terminal velocities emerged with inclusion of
comoving-frame line force \citep{cmf1}. Therefore, the discrepancy likely
results from line overlaps, which weaken the radiative force. However, unlike
the mass-loss rate, the wind terminal velocity is given by the solution of the
momentum equation from the stellar surface throughout the whole wind.
Consequently, the terminal velocity is sensitive to processes that appear on
large interval of radii and may modify the radiative force \citep{savi}. In
particular, wind X-ray emission and clumping may lead to higher radiative force
and, therefore, to higher wind terminal velocities \citep{nlteiii,irchuch}. This
can at least partially alleviate a disagreement between observational and
predicted values of terminal velocities for stars with $T_\text{eff}>30\,$kK.

As a result of ionization changes in the wind, the contribution of individual
elements strongly varies as a function of radius. Despite this, the radial
velocity profiles are smooth and do not show nonmonotonic change of the slope.
An exception is a model 150-40 at $Z_\odot/30$, which as a result of
\ion{Si}{iv} recombination shows more complex velocity profile that may resemble
double-beta velocity law \citep{hilmidoub,graham}.

\subsection{Line driving at low metallicities}

For the lowest metallicities and effective temperatures, the lines become
optically thin, marking the approach to the wind limit, where the homogeneous
winds cease to exist. This was studied by \citet{cnovit} in the context of pure
CNO-driven winds, who showed that in the domain of stars studied here the
minimum mass fraction of heavy elements required to drive a wind is about
$Z\approx10^{-4}$. As a result of the relatively low abundance of elements
heavier than oxygen, the model of purely CNO-driven wind provides a reasonable
approximation even for wind models at very low metallicities and solar mixture
of heavy elements. For instance, for the \ion{O}{v} 630\,\AA\ resonance line,
which dominates wind acceleration of the 400-5 wind model at lowest
metallicities, Eq.~(21) of \citet{cnovit} gives the minimum mass-loss rate
$7\times10^{-10}\,\msr$ for $Z=10^{-4}$; this value is not far off from the current theoretical
prediction in Table~\ref{ohvezpar}. This indicates that the model is very close
to the wind limit, where the lines cease to be optically thick and no longer
accelerate the wind.

For very low metallicities, $Z\leq0.1Z_\odot$, the wind is driven by just a few
strong lines (Fig.~\ref{trisil}). Therefore, the radiative force and the
mass-loss rates become sensitive to the fractional abundance of individual
elements. As a result, the mass-loss rate can be significantly affected by the
change of the composition caused by the mixing processes, which bring matter
affected by CNO burning up to the surface \citep{dusik}. In particular, oxygen,
which is a significant wind driver for the lowest metallicities and higher
effective temperatures, becomes depleted, leading to decrease in the wind
mass-loss rate. This can contribute to a significant scatter of the
observational mass-loss rates at the lowest metallicities
(Fig.~\ref{dmdttepportvx}).

Below the limit of chemically homogeneous winds (denoted as "no wind" in
Tables~\ref{bvele} and \ref{ohvezpar}), the radiative force can still influence
the structure of the atmosphere and trigger the chemical stratification
\citep{mirivi,tuhtalestra}. The wind limit shifts toward higher temperatures in
main-sequence stars with lower metallicities. Therefore, the stars with
chemically stratified atmospheres, so-called chemically peculiar stars, may
appear at higher temperatures in low-metallicity galaxies than in our Galaxy. In
the region of chemically peculiar stars, purely metallic winds might still be
possible \citep{babelb}.

\subsection{Inefficient shock cooling}
\label{wwp}

At low luminosities, the mass-loss rates estimated from observations are by one
to two orders of magnitude lower than the theoretical predictions
\citep{bourak,martin,linymarko}. This effect is possibly connected with wind
cooling in post-shock regions. The shocks appear in the wind likely as a result
of line-driven wind instability \citep{ocr,felpulpal} and the associated cooling
in the post-shock regions becomes inefficient in low-density environments
\citep{cobecru,nlteiii}. As a result of the inefficient cooling, a significant
fraction of the wind becomes hot, UV wind line profiles coming from ions with
low ionization energies get weakened and the line acceleration is reduced
\citep{predbehli}. This picture was supported by James Webb Space Telescope
mid-infrared (MIR) spectroscopy of the weak-wind star 10~Lac \citep{zakonici}, whose
forbidden lines of ions with high ionization energies indicate mass-loss rate
that is within the range of theoretical predictions.

From Rankine-Hugoniot jump conditions, the post-shock gas temperature can be
estimated as \citep[e.g.,][]{igor}
\begin{equation}
T_1=\frac{3}{16}\frac{\mu V_0^2}{k}=14\,\text{MK}\,
\zav{\frac{V_0}{1000\,\text{km}\,\text{s}^{-1}}}^2.
\end{equation}
Here, $V_0$ is the pre-shock velocity and $\mu$ is the mean atomic weight. With
the radiative losses per volume equivalent to $\epsilon=n_\text{H}^2\Lambda(T)$, where
$n_\text{H}$ is the hydrogen number density and $\Lambda(T)$ is the cooling
function, the characteristic time to radiate the thermal energy (cooling time)
can be roughly estimated as
\begin{equation}
\tau_\text{c}=\frac{\frac{3}{2}kT_1}{4n_\text{H}\Lambda(T_1)}.
\end{equation}
The factor of 4 in the denominator comes from the ratio of posts-shock and
pre-shock densities at an infinitely large shock, $\rho_1/\rho_0=4$
\citep{achimhydro}. A large fraction of wind material is occupied by very hot
gas in the case when the cooling time becomes comparable to the characteristic
wind time, $R_*/v_\infty$. With $\dot M=4\pi R_*^2m_\text{H}n_\text{H}v_\infty$,
this gives a limiting mass-loss rate of 
\begin{multline}
\dot M=\frac{3\pi}{2\Lambda} m_\text{H}kT_1R_*v_\infty^2=
1.2\times10^{-10}\,\msr\\\times
\zav{\frac{T_1}{10^6\,\text{K}}}\zav{\frac{R_*}{R_\odot}}
\zav{\frac{v_\infty}{1000\,\text{km}\,\text{s}^{-1}}}^2
\zav{\frac{\Lambda(T_1)}{10^{-22}\,\text{erg}\,\text{s}^{-1}\,\text{cm}^3}}^{-1}.
\end{multline}
For the calculation of the limiting mass-loss rate, we used the radiative losses from
\citet{jiste}, noting that they may be overestimated at around $10^5\,$K
\citep{odumdal}. For instance, for parameters corresponding to 10~Lac
\citep{moir} $R=4\,R_\odot$ and $v_\infty=1200\,\kms$, adopting $V_1=v_\infty/2$,
we have $T_1=5\,$MK.  For solar metallicity, this gives
$\Lambda(T_1)=2.7\times10^{-23}\,\text{erg}\,\text{s}^{-1}$ and a limiting
mass-loss rate of $1.3\times10^{-8}\,\msr$. This value is not far from the lower
limit of the 10~Lac mass-loss rate derived from infrared (IR) spectroscopy
$3\times10^{-8}\,\msr$ \citep{zakonici}. For the SMC metallicity,
$\Lambda(T_1)=1.0\times10^{-23}\,\text{erg}\,\text{s}^{-1}$ and the limiting
mass-loss rate is about $3\times10^{-8}\,\msr$ for the same terminal velocity.

For very low metallicities ($Z<Z_\odot/10)$ the cooling function becomes  lower by a
factor of about 2, shifting the limiting mass-loss rate to about
$10^{-7}\,\msr$. This implies that at the lowest metallicities considered here
basically all O stars should be affected by weak-wind problem. The problem is
less severe for B supergiants, which have slower winds and thus much
weaker shocks and shorter shock cooling times.

\subsection{Multicomponent effects}

At very low metallicities, the coupling between radiatively accelerated heavy
elements and light elements constituting the bulk of the wind, namely, hydrogen
and helium, can become inefficient. This leads to frictional heating or
decoupling of wind components \citep{treni,mcsw2,op}. The importance of these
effects can be estimated from the relative velocity difference between hydrogen
and individual heavy elements. 

We evaluated the relative velocity difference using Eq.~(17) of \citet{nlteii}.
For most of the models, the relative velocity difference is significantly lower
than 1, meaning that the multicomponent effects are negligible. However, for
several models with low mass-loss rates $\dot M\lesssim10^{-8}\,\msr$
(especially in those with higher wind velocities), the difference becomes higher
than one in the case of phosphorus in O stars and carbon, aluminium, and silicon
in B stars. While the decoupling of phosphorus with low abundance would perhaps
not significantly disrupt the flow, the decoupling of carbon can have more
significant consequences.

However, the relative velocity difference reaches values of a few at most.
Taking into account clumping, which is connected with higher densities in the
region where the wind is accelerated, the real velocity differences are likely
to be smaller. Thus, the effects of multicomponent structure are negligible in the
studied sample.

\section{Conclusions}

We provide new line-driven wind models for luminous OB stars with metallicities
down to $0.01\,Z_\odot$. The models were calculated with our global wind code
METUJE, which solves the hydrodynamical equations from nearly hydrostatic
photosphere to supersonically expanding stellar wind. The radiative force
accelerating the wind was derived from the solution of the radiative transfer
equation in the comoving frame. The corresponding ionization and excitation
structure was derived from the kinetic equilibrium equations (NLTE equations).

The models predict the wind structure including the wind mass-loss rate and
terminal velocity just from the basic stellar parameters; namely, the stellar
effective temperature, radius, mass, and chemical composition. The wind
mass-loss rate depends mostly on the stellar luminosity and decreases with
decreasing metallicity roughly as $\dot M\sim Z^{\alpha}$. For the metallicities
down to the metallicity corresponding to the Magellanic Clouds ($Z\geq 0.2\,Z_\odot$),
the metallicity dependence of the mass-loss rate is rather weak, at
$\alpha\approx0.6$. For lower metallicities, this relationship steepens and
$\alpha$ becomes higher than 1.

The wind mass-loss rate also significantly depends on the stellar effective
temperature. The most significant feature coming from the temperature dependence
of the wind mass-loss rate is the increase in the mass-loss rate with decreasing
temperature for B supergiants with $T<20\,$kK, termed the bistability jump. This
increase originates due to iron recombination and vanishes for subMagellanic
metallicities ($Z\leq0.1\,Z_\odot$) coming from the weak iron contribution to
the radiative force for very low metallicities.

Down to metallicities corresponding to the Magellanic Clouds, the predicted
mass-loss rates  agree reasonably well with observational estimates. On average, our
mass-loss rates nicely reproduce the observational values with an exception of
the SMC O stars. However, for the SMC metallicity our models reasonably
reproduce the mass-loss rates in B supergiants, which other available
predictions tend to either over- or underpredict to a significant degree. Moreover, the
disagreement between observational mass-loss rate estimates and theoretical
mass-loss rate predictions only concerns SMC O stars with low mass-loss rates.
Therefore, the discrepancy is most likely caused by the inefficient shock
cooling (weak-wind problem) and not by inadequate predictions.

For subMagellanic metallicities, both the theoretical mass-loss rate predictions
and the observational values display significant scatter. On the theoretical
side, the scatter is caused by a low number of lines significantly accelerating
the wind, which leads to an increased sensitivity in the radiative acceleration
to the detailed ionization structure of the wind. The observational values for
low mass-loss rates $\dot M\lesssim10^{-7}\,\msr$ and subMagellanic
metallicities become affected by the inefficient shock cooling in the wind. The
inefficient cooling leaves significant fraction of the stellar wind at very high
temperatures weakening classical signatures of line-driven wind and the
radiative force. Limiting the comparison only to stars with sufficiently high
mass-loss rates, where the shock cooling is efficient, we arrive at good
agreement between observational and theoretical mass-loss rate estimates.

The wind terminal velocity scales with the escape speed and with the stellar
effective temperature. The predicted terminal velocities  correspond well to
the observational values up to the effective temperature of $30\,$kK. Above this
value the models underpredict the observational values likely as a result of
neglected wind clumping and X-ray emission. Despite this shortcoming, an
extrapolation of terminal velocities derived below $30\,$kK to higher effective
temperatures provides estimates that nicely agree with observations.

Here, we provide fits to theoretical predictions of the wind mass-loss rate and
terminal velocity. The fits can be used in the evolutionary calculations and for
the modeling of interaction of hot stars with their surrounding circumstellar
environment.

\section*{Data availability} 

Wind model output files are available in electronic form at the CDS via
anonymous ftp to cdsarc.u-strasbg.fr (130.79.128.5) or via
http://cdsweb.u-strasbg.fr/cgi-bin/qcat?J/A+A/.

\begin{acknowledgements}
This work was supported by grant GA \v{C}R 25-15910S. Computational resources
were provided by the e-INFRA CZ project (ID:90254), supported by the Ministry of
Education, Youth and Sports of the Czech Republic. The Astronomical Institute
Ond\v{r}ejov is supported by a project RVO:67985815 of the Academy of Sciences
of the Czech Republic.
\end{acknowledgements}

\bibliographystyle{aa}
\bibliography{papers}

\end{document}